\documentclass[aip,apl,amsmath,amssymb,floatfix,preprint,letter]{revtex4-1}
\usepackage{graphicx}
\usepackage{dcolumn}
\usepackage{bm}

\begin{document}

\title{Coherent-feedback control strategy to suppress spontaneous switching in ultra-low power optical bistability}

\author{Hideo Mabuchi}\email{hmabuchi@stanford.edu}
\affiliation{Edward L.\ Ginzton Laboratory, Stanford University, Stanford, CA 94305, USA}

\date{15 January 2011}

\begin{abstract}
An optical resonator with intracavity Kerr nonlinearity can exhibit dispersive bistability suitable for all-optical switching. With nanophotonic elements it may be possible to achieve attojoule switching energies, which would be very attractive for ultra-low power operation but potentially problematic because of quantum fluctuation-induced spontaneous switching. In this Letter I derive a quantum-optical model of two Kerr-nonlinear ring resonators connected in a coherent feedback loop, and show via numerical simulation that a properly designed `controller' cavity can significantly reduce the spontaneous switching rate of a bistable `plant' cavity in a completely embedded and autonomous manner.
\end{abstract}

\maketitle

\begin{figure}[b!]
\includegraphics[width=0.55\textwidth]{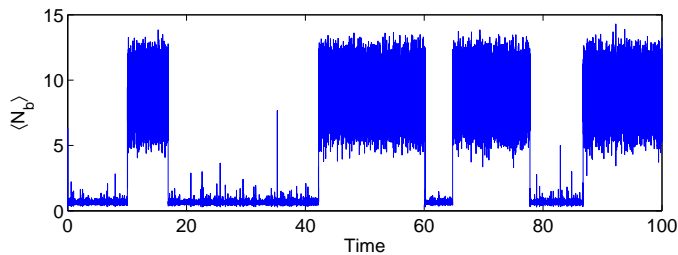}
\vspace{-0.1in}
\caption{\label{fig:qt} Quantum trajectory simulation of the mean intracavity photon number for a driven optical resonator including a Kerr nonlinear medium. The jumps between low and high photon-number states, which occur spontaneously with the drive amplitude held fixed, are symptomatic of quantum destabilization of dispersive optical bistability.}
\vspace{-0.1in}
\end{figure}

\noindent Although current nanophotonics research focuses mainly on the design and demonstration of individual optical components, future progress towards technological relevance will surely require the development of nanophotonic {\em circuit} theory at a level of sophistication comparable to that of modern electronics. As the performance regime of interest for nanophotonic technologies extends to picosecond switching times and attojoule (few-photon) switching energies, quantum-optical effects will be of great practical significance even if the information processing paradigm remains purely classical ({\it i.e.}, before the advent of true quantum information technology). Rigorous yet user-friendly theoretical methods (based on new generalizations of classical stochastic systems theory~\cite{Goug09}) for the quantum-optical analysis of photonic circuits have recently been developed~\cite{Kerc10}, but compatible algorithmic design methods~\cite{Jame08a,Nurd09} are still quite limited in scope. It is thus an opportune moment to begin investigating relatively simple nanophotonic circuit `motifs' in order to exercise our new analysis methods and to provide guidance for subsequent work on more complex component networks. Given the ubiquity of feedback configurations for noise suppression within microelectronic circuits, it seems natural to focus such preliminary exploration on coherent (optical) feedback motifs for managing quantum fluctuations in ultra-low power nanophotonic circuits.

\begin{figure}[b!]
\includegraphics[width=0.55\textwidth]{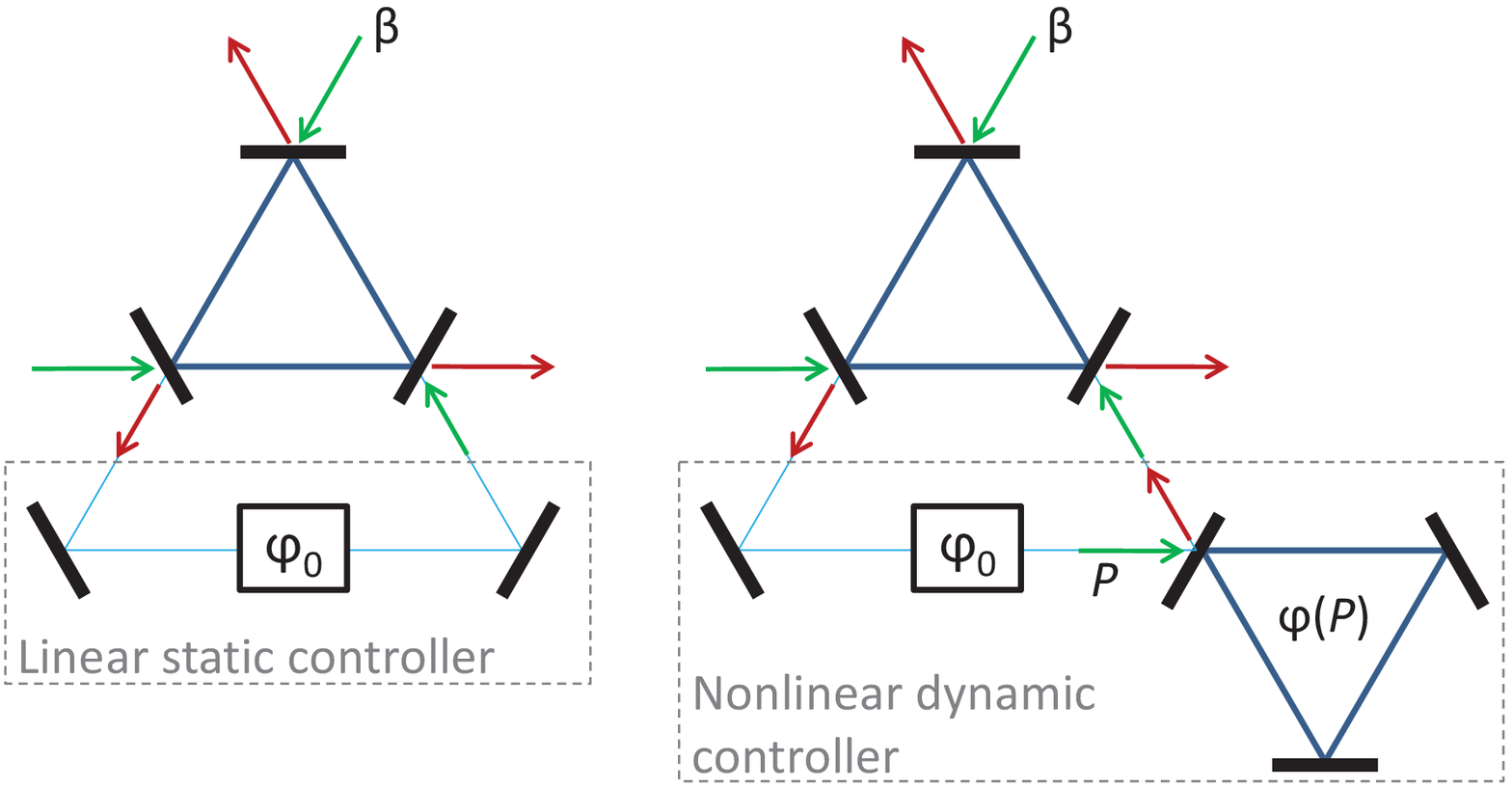}
\includegraphics[width=0.55\textwidth]{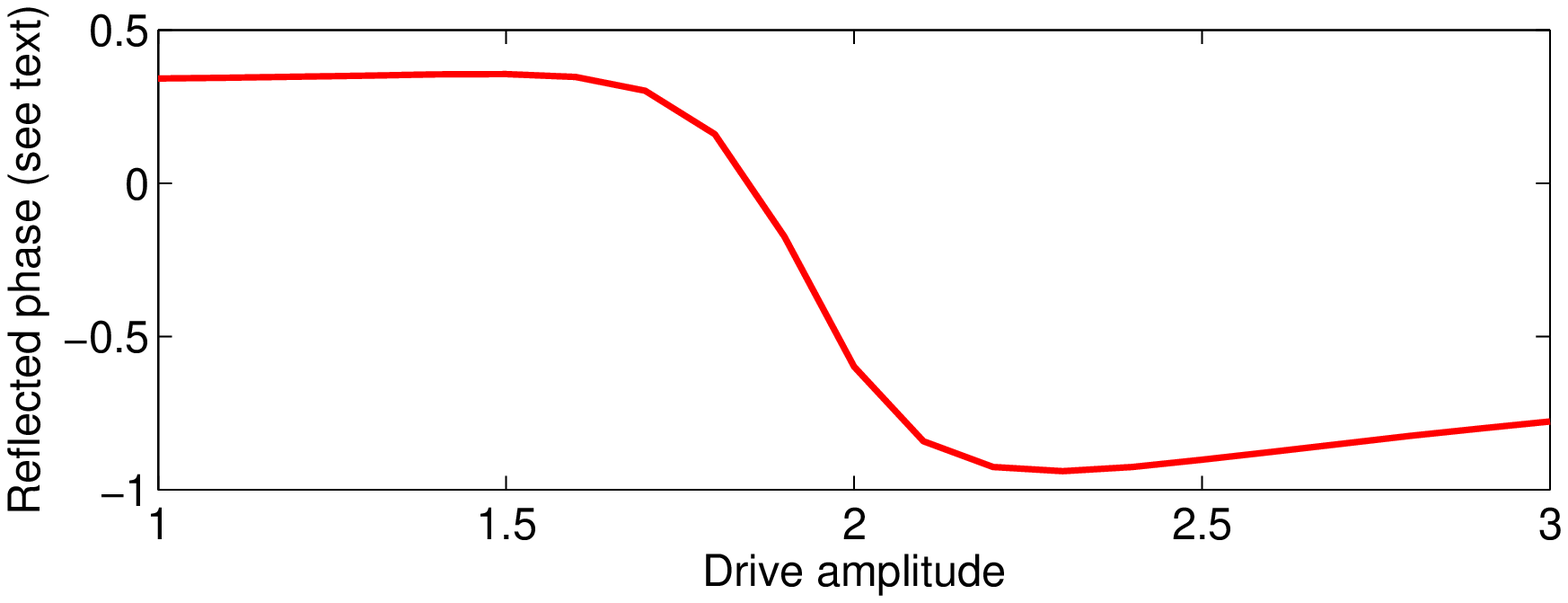}
\vspace{-0.1in}
\caption{\label{fig:schemats} Upper panels: Schematic diagrams of coherent feedback configurations with linear static (left) and nonlinear dynamic (right) controllers. Lower panel: Nonlinear steady-state phase shift of a coherent field upon reflection from the Kerr-nonlinear controller cavity (see text for details).}
\vspace{-0.1in}
\end{figure}

Here we consider a coherent feedback~\cite{Jame08a,Nurd09,Mabu08a} strategy for suppressing spontaneous switching in dispersive optical bistability. Dispersive bistability is of interest as a potential physical basis for the design of ultra-low power nanophotonic switches~\cite{Yani03,Yang07,Noza10,Kuma10}, but in the attojoule switching regime where the logical states are separated by a small number of photons, quantum fluctuations will induce unwanted spontaneous switching~\cite{Arme06,Arme09,Kerc11b} that must be accounted for in circuit design. For example, Fig.~\ref{fig:qt} shows a simple quantum trajectory simulation (performed using the Quantum Optics Toolbox for Matlab~\cite{Tan99}) of the mean intracavity photon number for a Kerr-nonlinear optical resonator, assuming parameters (cavity decay rate $\kappa_b=150$, drive detuning $\Delta_b=5\kappa_b$, and nonlinear coefficient $\chi_B=-\Delta_b/10\sqrt{2}$) that classically would be expected to support dispersive optical bistability with attojoule separation between the logical high and low states. The quantum model~\cite{WM}, corresponding to the Master Equation $(\hbar=1)$
\begin{eqnarray}
\label{eq:MEOL}\dot{\rho}&=&-i[H_b,\rho]+\kappa_b\left\{b\rho b^\dag - \frac{1}{2}b^\dag b\rho - \frac{1}{2}\rho b^\dag b\right\},\\
\label{eq:Hb}H_b&=&\Delta_b b^\dag b + \chi_b b^\dag b^\dag b b + i\sqrt{\kappa_{b3}}(\beta^* b - \beta b^\dag),
\end{eqnarray}
clearly predicts spontaneous transitions that would compromise the performance of such a device in a photonic switching context. Here $b$ is the annihilation operator for the intracavity field mode, $\beta$ is the complex amplitude of a coherent drive field, and $\kappa_{b3}\le\kappa_b$ is the partial decay rate associated with the input coupler of the resonator. In Fig.~\ref{fig:qt} we have set $\beta\sqrt{\kappa_{b3}}=10.4934\sqrt{50}$ to achieve approximately equal time-average occupation of the low- and high-photon number states.

In order to motivate our coherent-feedback stabilization strategy for suppression of such `quantum jumps' we first consider a feedback configuration with a linear static controller. If we assume that the bistable (`plant') cavity has three distinct input-output ports corresponding to the bias input and the feedback-loop input and output, we can model~\cite{SM} the effects of a simple optical feedback loop with unit gain and total phase shift $\varphi$ (as depicted in the upper left panel of Fig.~\ref{fig:schemats}) using the Master Equation~(\ref{eq:MEOL}) with $H_b\rightarrow H_b(\varphi)$ and $\kappa_b\rightarrow\kappa_b(\varphi)$, where
\begin{eqnarray}
H_b(\varphi)&=&H_b + \sin(\varphi)\sqrt{\kappa_{b1}\kappa_{b2}}b^\dag b,\\
\kappa_b(\varphi)&=&\kappa_{b3}+\vert\sqrt{\kappa_{b1}}+e^{i\varphi}\sqrt{\kappa_{b2}}\vert^2.
\end{eqnarray}
Here $\kappa_{b1,2}$ are the partial decay rates associated with coupling to the feedback loop; it is assumed that $\kappa_{b1}+\kappa_{b2}+\kappa_{b3}=\kappa_b$ and in what follows we will set them equal. It can be seen that the net effects of the feedback loop are a $\varphi$-dependent frequency pulling of the effective drive detuning and a $\varphi$-dependent change in the effective cavity decay rate. Either or both of these effects could potentially be used to suppress spontaneous switching of the bistable cavity if $\varphi$ could be adjusted to a value that stabilizes the low state when the state is low, and to a value that stabilizes the high state when the state is high.

To realize the desired form of nonlinear dynamic controller we consider an auxiliary (`controller') Kerr-nonlinear optical cavity (with parameters $\kappa_a=50$, $\Delta_a=3\kappa_a$, $\chi_a=-\Delta_a/8$) connected as shown in the upper right panel of Fig.~\ref{fig:schemats}. Such a cavity imparts a drive amplitude-dependent phase shift on the beam that reflects from its input coupler. The lower panel of Fig.~\ref{fig:schemats} shows an approximate representation of reflected phase versus drive amplitude (in units that would correspond to $\vert\langle b\rangle\vert$ for a coherent state of the plant), computed using a Master Equation analogous to~(\ref{eq:MEOL}) (the approximation consists in considering the complex phase of $\langle a\rangle$ as the phase of the intracavity field, where $a$ is the corresponding annihilation operator, but this introduces small errors as the intracavity field becomes somewhat non-Gaussian at high drive amplitude). Our basic coherent control strategy is to choose $\varphi$ such that the overall feedback phase (including the amplitude-dependent phase shift contributed by the controller cavity) is close to $\pi$ when the plant is in the low photon-number state (decreasing $\kappa_b(\varphi)$ and thus reducing the efficiency of pumping by the detuned bias input $\beta$), and closer to zero when the plant is in the high photon-number state.

The Master Equation for this coherent feedback configuration is given by~\cite{SM}
\begin{eqnarray}
\nonumber\dot{\rho}&=&-i[H,\rho]+\sum_{j=1,2}\left\{L_j\rho L_j^\dag - \frac{1}{2}L_j^\dag L_j\rho - \frac{1}{2}\rho L_j^\dag L_j\right\},\\
\nonumber H&=&H_a+H_b(\varphi)+\frac{\sqrt{\kappa_{a}\kappa_{b2}}}{2i}(e^{i\varphi}a^\dag b-e^{-i\varphi}ab^\dag)\\
\nonumber &&+\frac{\sqrt{\kappa_{a}\kappa_{b1}}}{2i}(ab^\dag-a^\dag b),\\
\nonumber L_1&=&\sqrt{\kappa_{a}}a+(e^{i\varphi}\sqrt{\kappa_{b2}}+\sqrt{\kappa_{b1}})b,\\
\label{eq:MECL}L_2&=&\sqrt{\kappa_{b3}}b,
\end{eqnarray}
where $H_a$ is obtained by setting $b\rightarrow a$ in Eq.~(\ref{eq:Hb}). Given the highly nonlinear nature of the model, a simple numerical procedure was used to find a value of $\varphi$ that resulted in reduced spontaneous transition rate. This was done by holding all other parameters fixed and determining the steady-state density matrix for various trial values of $\varphi$; it was found that plant bistability is recovered in the closed-loop configuration only in narrow ranges of $\varphi$ around $\varphi\approx 2.3681$ and $\varphi\approx 5.2277$. The former setting successfully reduces the spontaneous transition rate in the plant cavity, while the latter setting increases it.

\begin{figure}[tb!]
\includegraphics[width=0.55\textwidth]{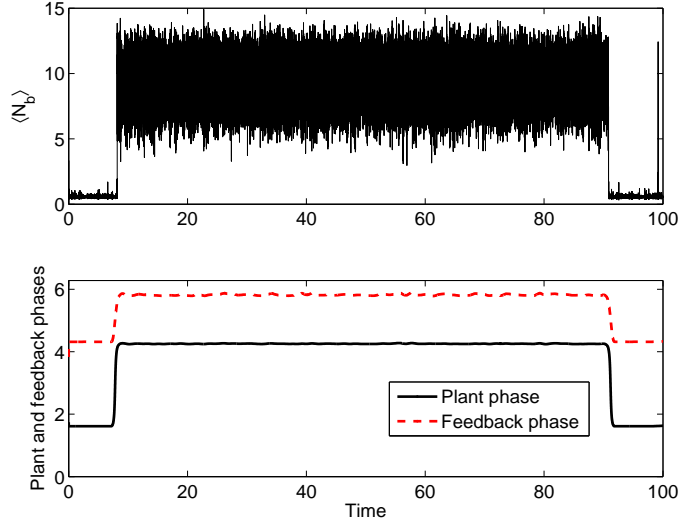}
\vspace{-0.1in}
\caption{\label{fig:clmc} Upper plot: Quantum trajectory simulation of the mean plant intracavity photon number assuming nonlinear dynamic coherent feedback control with $\varphi=2.3681$. Lower plot: Plant and feedback phases (smoothed, see text) displayed on the same time axis as the upper plot.}
\vspace{-0.1in}
\end{figure}

The upper panel of Fig.~\ref{fig:clmc} displays a quantum trajectory simulation of the closed-loop model with $\varphi=2.3681$; spontaneous transitions between low and high photon-number states are clearly still present but occur at a reduced rate. The lower panel of Fig.~\ref{fig:clmc} shows the complex phase of $\langle b\rangle$ (corresponding approximately to the plant phase) and the phase of the coherent feedback field after reflection from the controller cavity (computed as the complex phase of $\sqrt{\kappa_{a}}\langle a\rangle+e^{i\varphi}\sqrt{\kappa_{b2}}\langle b\rangle$) on the same time axis (although significantly low-pass filtered to reduce shot-noise fluctuations). The difference of these two values corresponds to the phase shift of the coherent feedback loop. In accordance with the intuitive strategy described above, it can be seen that the feedback phase shift takes a value ($\approx 2.7$) that decreases the effective plant decay rate (by a factor $\approx 0.4$) when the plant is in the low photon-number state. In the high photon-number state the feedback phase shift fluctuates around a value $\approx 1.57$.

\begin{figure}[tb!]
\includegraphics[width=0.55\textwidth]{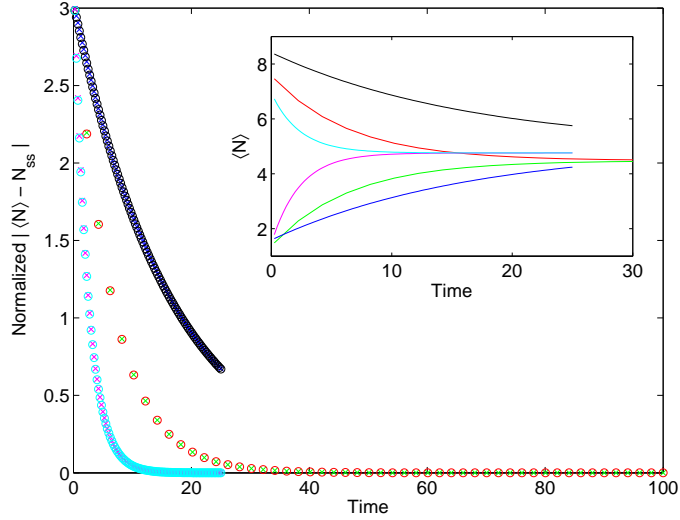}
\vspace{-0.1in}
\caption{\label{fig:decays} Inset: Mean plant photon number as a function of time for two different initial conditions and in the open-loop (green/red), closed-loop $\varphi=2.6381$ (blue/black), and closed-loop $\varphi=5.2277$ (magenta/cyan) cases. Main plot: Normalized versions of the data from the inset (see text).}
\vspace{-0.1in}
\end{figure}

The inset of Fig.~\ref{fig:decays} displays the evolution of the mean plant photon number $\langle b^\dag b\rangle$ as a function of time, starting from either a zero- or nine-photon initial condition (computed in the closed-loop cases by numerically integrating the master equation and in the open-loop case by numerical diagonalization of the Liouville super-operator~\cite{Tan99}). In the long time limit both initial conditions regress to a steady-state value, which is the average of the conditional mean photon number in the low and high logical states. Results are shown for the open-loop case and the dynamic coherent feedback cases with $\varphi=2.3681$ (longest decay timescale) and $\varphi=5.2277$ (shortest decay timescale). It can be seen that the low and high logical states, and therefore the steady-state photon number, vary somewhat among the three cases. The main plot of Fig.~\ref{fig:decays} displays the same data in a normalized fashion. The curves correspond to $\vert\langle b^\dag b\rangle(t) - \langle b^\dag b\rangle(\infty)\vert$ and are normalized to have identical values at the first time point (chosen as $t=0.25$ in order to omit initial transients while the simplistic initial conditions equilibrate to their nearby logical states). The fact that the curves obtained with low ($\times$) and high ($\circ$) initial conditions coincide shows that low$\rightarrow$high and high$\rightarrow$low transitions are suppressed (or enhanced) equally. The regression timescale in the $\varphi=2.3681$ closed-loop case is longer than that of the open-loop case by a factor $\approx 2.5$.

It should be noted that within the two-cavity coherent feedback configuration we have considered, and with the key `structural' parameters ($\chi_b$, $\kappa_b$) of the plant cavity held fixed, there remains a great deal of room for optimizing the operating conditions ($\beta$, $\Delta_b$) and controller parameters ($\Delta_a$, $\kappa_a$, $\chi_a$ and $\varphi$) to achieve potentially superior suppression. Given the rather demanding nature of the numerical computations involved (a total Hilbert space dimension of 625 was used in this work and Master Equation integrations were essential), a brute-force scan of so many degrees of freedom would not seem feasible but it seems likely that a more principled computational optimization approach could be developed.

Recent theoretical investigations---based on classical electromagnetic models---of circuit motifs~\cite{Li10} and optimal pulse shaping for switching applications~\cite{Sand10} have offered a glimpse of the great potential for innovative engineering at the signals-and-systems (as opposed to device physics) level in nanophotonics. Here we have attempted to extend this exploration to the quantum optical regime of attojoule switching energy, demonstrating that new theoretical methods~\cite{Goug09} can be used to analyze intuitive coherent feedback control schemes in quantitative detail.

This research was supported by the ARO (W911NF-08-1-0427) and by DARPA-MTO (FA8650-10-1-7007).

\end{document}